\documentclass[usenatbib,usegraphicx,referee]{mn2e}

\def\araa{ARA\&A}
\def\mnras{MNRAS}
\def\apj{ApJ}
\def\aj{AJ}
\def\apjl{ApJL}
\def\aap{AAP}

\def\na{NewA}

\def\P{{\cal P}}

\def\HI{{\rm H~{\sc i}~}}

\usepackage{epsfig}
\usepackage{graphicx}
\begin{document}
\title{Effect of turbulent velocity on the \HI intensity fluctuation power spectrum from spiral galaxies}
\author[Prasun Dutta]
{Prasun Dutta$^{1}$\thanks{Email:prasun@iiserb.ac.in},  
\\$^{1}$ IISER Bhopal, ITI Gas Rahat Building, Bhopal-462023, India.}

\date{}
\maketitle
\begin{abstract}
We use numerical simulations to  investigate effect of  turbulent velocity on the power spectrum of \HI intensity  from external galaxies when (a) all emission is considered, (b) emission with velocity range smaller than the turbulent velocity dispersion is considered. We found that for case (a) the intensity fluctuation depends directly only on the power spectrum of the column density, whereas for case (b) it depends only on the turbulent velocity fluctuation.  We discuss the implications of this result in real observations of \HI fluctuations.
\end{abstract}


\section{Introduction}
Observationally, power spectrum of the \HI intensity fluctuation in our Galaxy suggest existence of scale invariant structures in the \HI density over  length scales ranging  as wide as sub parsec to a few hundred parsec \citep{1983A&A...122..282C, 1993MNRAS.262..327G}. These structures are understood \citep{2004ARA&A..42..211E} in terms of compressible fluid turbulence in the interstellar medium (ISM).  In present theoretical understanding of ISM dynamics, compressible fluid turbulence plays an important role in the ISM evolution, energy transfer, star formation etc. Possible source of energy into the turbulence cascade is however debated, though is mostly ascribed to the  
supernova shocks as a large scale energy input. Different techniques have been developed to measure the velocity spectrum of the turbulence and hence infer the energy involved in the process. Techniques, originally developed to get the velocity structures in the Galaxy, 
include statistics of centroid of velocities \citep{2009RMxAC..36...45E}, velocity coordinate spectrum or VCS \citep{2009RMxAC..36...54P}, velocity channel analysis or VCA (\cite{2000ApJ...537..720L}, henceforth LP00), spectral correlation function \citep{2001ApJ...555L..33P} etc. We particularly bring  attention of our reader here to VCA, which has been applied \citep{2010ApJ...714.1398C, 2009ApJ...693.1074C, 2008ApJ...688.1021C, 1988A&A...191...10S} for observations in our Galaxy as well as nearby dwarf galaxies like Large and Small Magellanic Clouds. It was found that the velocity fluctuations also follow a power law. Interested reader may have a look at the article LP00 for a complete description of VCA, here we outline the basic principle behind this analysis. VCA aims to extract the velocity power spectrum by comparing the power spectrum of intensity averaged over the entire velocity range of observation with the same averaged over a relatively small velocity range, smaller than the expected turbulence velocity dispersion. Differential rotation of our Galaxy allows us to have a direct mapping of the velocity values in the position-position-velocity data cube with the line of sight distance to the observing cloud. This in turn let us estimate the three dimensional power spectrum of the \HI density fluctuations from the Galaxy. However,  turbulence velocity fluctuations  change  the velocity to distance mapping and hence also modifies the intensity power spectrum. This is precisely what VCA explores. 

\citet{2006MNRAS.372L..33B} have used a visibility based power spectrum estimator to measure the intensity fluctuation power spectrum of the nearby dwarf galaxy DDO~210. They infer that the density power spectrum has a slope of $-2.75$ over a length scale range of $80$ to $500$ pc. They used the VCA  with their position-position-velocity data cube and inferred an upper limit to  the  slope of the velocity power  spectrum. \citet{2008MNRAS.384L..34D, 2009MNRAS.398..887D, 2009MNRAS.397L..60D} has extended these study to several external dwarf and spiral galaxies and have estimated the density power spectrum.  Recently, \citet{2013NewA...19...89D, 2013MNRAS.436L..49D} has estimated the power spectrum of the $18$ spiral galaxies from THINGS \footnote{THINGS: The \HI Nearby Galaxy Survey \citep{2008AJ....136.2563W}.} sample and found that the power spectrum of column density follow a power law over the length scales ranging $400$ pc to $16$ kpc considering the entire sample. Slope of the power spectra for most of these galaxies was found to be within $-1.5$ to $-1.8$. Generating mechanism of these large scale structures are yet to be understood. Measuring statistics of the velocity  fluctuations would help us understand the dynamical phenomena responsible for these structures. 

We note here two main difference between the position-position-velocity data cubes of the \HI emission observation in the Galaxy and external spiral galaxies. Line of sight to the observations for the \HI emission from our Galaxy is mostly along the plane of the disk, while, the external galaxies are mostly for the face on galaxies and the line of sight is perpendicular to the disk. For observations in our Galaxy, different velocity slices of the data cube can be considered to be at different distances but at the same angular direction in the sky, whereas,  for the external galaxies,  different velocity slices of the data cube  originates from different parts of the galaxy. This suggests that it would not be wise to directly use the results obtained in LP00 while inferring observations from external galaxies.

As there exist no direct position to velocity mapping for the external spiral galaxies, an analytical investigation on the effect of the turbulence velocity is not straight forward, we refer to the numerical methods here. In this letter we perform numerical simulation to access  how the intensity power spectrum is modified with the velocity fluctuations for the spiral galaxies.  Section (2) gives a brief outline to our approach and the section (3) describes the numerical investigation we have performed. Results and discussions are discussed in section (4). We conclude in section (5).

\section{Modelling \HI emission from spiral galaxy}
We adopt a coordinate system centred at the \HI cloud in concern (or the external galaxy) with the line of sight direction aligned to the $z$ axis, such that
\begin{equation}
\vec{r}\ =\ (x, y, z)\ =\ (\vec{R}, z),
\end{equation}
where $\vec{R} = (x,y)$ is a two dimensional vector in the sky plane. At small optical depth limit, the specific intensity of radiation \citep{2011piim.book.....D} with rest frequency $\nu_{0}$ originated from a gas at $\vec{r}$ having temperature $T$ is given as
\begin{equation}
I (\vec{R}, v)\ =\ I_{0}\,  \int d\,  z\, n_{HI} (\vec{r})\, \phi (v),
\end{equation}
where $v = c ( \nu_{0} - \nu)$,  $\nu$ is the frequency of observation,  $I_{0} = \frac{3 h \nu_{0} A_{21}}{16 \pi}$ and $\phi(v)$ is the line shape function:
\begin{equation}
\phi(v) = \phi_{0}\, \exp \left [ - \frac{ \left ( v - v_{z} (\vec{r}) \right ) ^{2} } {2 \sigma^{2}(\vec{r}) } \right ].
\end{equation}
Here $v_{z}(\vec{r})$ is the line of sight component of the velocity of the gas and $\sigma(\vec{r}) = \sqrt{\frac{ k_{b} T}{m_{HI}}}$ \footnote{$k_{b}$ : Boltzmann constant, $m_{HI}$ : mass of hydrogen atom} is the thermal velocity dispersion. In practice, the observed specific intensity is always averaged over a velocity width $\delta v$ around $v$, hence
\begin{equation}
I ^{obs}(\vec{R}, v, \delta v)\ =\ \frac{1}{\delta v} \int _{v - \delta v/2}^{v + \delta v/2} dv' I (\vec{R}, v').
\end{equation}
Clearly,
\begin{equation}
\lim_{\delta v \to \infty}\ I ^{obs}(\vec{R}, v, \delta v)\ =\ I_{0}\, N_{HI}(\vec{R}),
\end{equation}
where, $N_{HI}(\vec{R}) = \int dz\, n_{HI}(\vec{r})$ is the column density. In practice, as the emission from the galaxy falls of to zero beyond a certain velocity, say $\pm  \Delta v$, it is sufficient to carry the integration in the above equation over the range $-\Delta v$ to $\Delta v$. Here we consider that the galaxy has no overall motion.

Compressible fluid turbulence in the ISM of the galaxies induce scale invariant fluctuations in the density as well as velocity.  Power spectrum of the column density fluctuation is given as
\begin{equation}
P_{N_{HI}}(K) \ =\ \int \ d\vec{X}\, e^{-i \vec{K} . \vec{X}} \langle  N_{HI}(\vec{R}+ \vec{X})  N_{HI}(\vec{R})\rangle,
\end{equation}
where the averaging is performed over all possible values of $\vec{R}$ and all directions assuming homogeneity and isotropy in the random fluctuations. We define the power spectrum of the observed intensity fluctuation as 
\begin{equation}
P(K , \delta v) =\int d\vec{X}\, e^{-i \vec{K} . \vec{X}} \langle  I ^{obs}(\vec{R}+ \vec{X}, v, \delta v)  I ^{obs}(\vec{R}, v, \delta v)\rangle
\end{equation}
Here we have  assumed the homogeneity and isotropy of the intensity field and that the  intensity power spectrum  is independent of the centroid of the velocity $v$. Clearly,
\begin{equation}
\lim_{\delta v \to \Delta v} P(K , \delta v) \ \propto P_{N_{HI}}(K).
\end{equation}
This has been used extensively in literature to estimate the \HI column density power spectrum of our Galaxy \citep{1983A&A...122..282C,1993MNRAS.262..327G}, external dwarf \citep{2006MNRAS.372L..33B, 2009MNRAS.398..887D} and spiral galaxies \citep{2013NewA...19...89D}. The power spectra is found to follow power laws indicating turbulence to be operational, hence
\begin{equation}
P_{N_{HI}}(K) \ =\ A_{N_{HI}} K^{\alpha}
\end{equation}

The line of sight component of velocity $v_{z}(\vec{r})$ has component from the systematic rotation of the galaxy $v^{\Omega}(\vec{r})$, as well as form the random motion of the cloud because of turbulence $v^{T}(\vec{r})$, i.e, $v_{z}(\vec{r}) = v^{\Omega}(\vec{r}) + v^{T}(\vec{r})$. Power spectrum of the turbulent velocity component is also expected to follow power law
\begin{equation}
P_{v^{T}}(K) \ =\ A_{v_{T}} K^{\beta}.
\end{equation}
LP00 has investigated the nature of $P(K, \delta v)$ in detail in order to estimate the modification of the \HI power spectrum by turbulence. They show that for observed \HI gas in our galaxy,  for $\alpha > -3$, 
\begin{equation}
\lim_{\delta v < \sigma_{T}} P(K, \delta v) \propto K^{\alpha + \beta/2},
\end{equation}
while as we expect,
\begin{equation}
\lim_{\delta v \to \sigma_{T}} P(K, \delta v) \propto K^{\alpha} \propto P_{N_{HI}} (K).
\end{equation}
Hence, by estimating the power spectra in two different limits above one can infer the velocity power spectrum slope. This method, usually known as velocity channel analysis, has been used to estimate the velocity fluctuation power spectrum of \HI in our Galaxy and nearby dwarf galaxies.

It is important to realise that the particular direct linear mapping between $v_{z}$ and $z$ was exploited in VCS, is rather different when we consider \HI emission from an external galaxy. Considering tilted ring model, in later case,  $v^{\Omega}(\vec{r})$ depends on the galacto-centric radius, position and inclination angles. Moreover, at a given $v$ with $\delta v < \Delta v$, only a part of the galaxy's disk is visible. In this letter we attempt to see how the turbulent velocity modifies the \HI power spectrum for the external spiral galaxies and investigate if a similar procedure as VCS can be adopted to estimate the velocity fluctuation spectrum. 

\subsection{Simplifications}
In order to simulate the \HI emission from the external galaxies we adopt the following simplifications. Note that in this work we are not interested to simulate all aspect of the \HI emission from the external galaxies, rather we are interested in investigating the modification in the power spectrum due to the turbulent velocity, which justifies these simplifications.
\begin{itemize}
\item In case of a spiral galaxy the average \HI profile $W(\vec{r})$ varies with the galacto-centric radius as well as in vertical direction. This leads to modification in the \HI power spectrum as discussed in \citet{2009MNRAS.398..887D}. Here, we consider $W(\vec{r})$ to be independent of $\vec{r}$. It is to be noted that this simplification also means that we are assuming the galaxy's disk to be thick. We shall discuss the effect of this in the conclusions section.
\item
Systematic rotation of the galaxy $v^{\Omega}(\vec{r})$, depends on the inclination and position angle as well as galacto-centric radius. To simplify matter we assume here that the position angle and inclination angle $(i)$ of the galaxy do not change with galacto-centric radius and adopt a flat rotation curve with tangential velocity $v_{0}$. In such a case, we can write
\begin{equation}
v^{\Omega}(\vec{r}) \ =\ \frac{v_{0} sin(i) x}{\sqrt{x^2 + y^2 cos^{2}(i)}}
\end{equation}
\item
ISM is known to be in pressure equilibrium (see \cite{2003ApJ...587..278W} and references therein) and more than one temperature gas coexists in it. This  means, in principle,  we need to consider different temperatures at different part of the galaxy and hence a varying $\sigma(\vec{r})$. This would give rise to an additional fluctuation in the observed specific intensity. Here we assume that the gas across the galaxy is at a constant temperature and we adopt the temperature that of the cold gas.
\end{itemize}
\section{Simulation}

\begin{figure}
\begin{center}
\epsfig{file=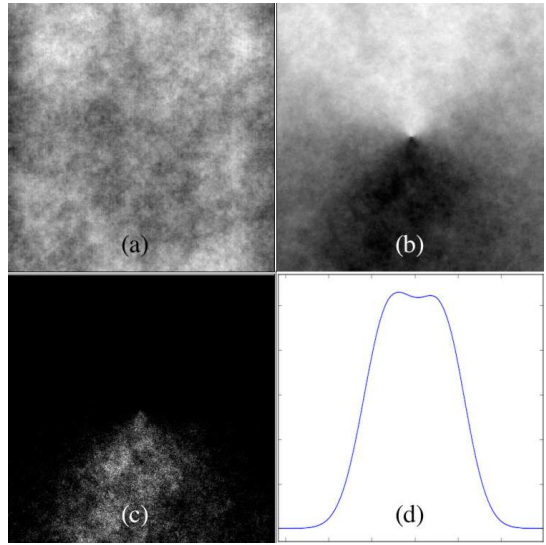,width=2.8in}
\end{center}
\caption{Plot showing map of  (a) Column density, (b) line of sight velocity, (c) observed specific intensity for $\delta v = \sigma_{T}/2$. Part (d) is the  integrated line profile of the simulated galaxy with $\alpha = -2.5$, $\beta = -2.5$, $i = 10^{\circ}$.}
\label{fig:NHI}
\end{figure}

We divide the simulation volume into $NGrid^{3}$ individual grids (cubes) which represents individual \HI cloud with associated $n_{HI}$ and $v_{z}$. As discussed in the previous section, we have kept $\sigma$ constant everywhere across the simulation volume. It can be shown that for a thick cube the power spectrum of the three dimensional density distribution has the same slope of that of the projected density, like the column density here (\citet{2009MNRAS.398..887D}, LP 00). Hence, we generate $\delta n_{HI}$ and $v^{T}$ such that they follow Gaussian distribution with power spectrum of slope $\alpha$ and $\beta$ respectively. \citet{2013MNRAS.436L..49D} have estimated the amplitude of the \HI fluctuations for six galaxies of the THINGS sample \citep{2008AJ....136.2563W}. They found that the amplitude of the column density fluctuations are approximately $1/10^{th}$ that of the mean column density for the galaxies. Hence, here we consider
\begin{equation}
n_{HI} \ =\ n_{0} \left [ 1 + f_{n_{HI}} \delta n_{HI} (\vec{r}) \right ],
\end{equation}
with $f_{n_{HI}} = 0.1$ and  $\delta n_{HI}(\vec{r})$, the  random component because of turbulence.  \citet{2009AJ....137.4424T} has estimated the \HI velocity dispersion for the galaxies in the THINGS from  Moment-II maps and found that the  turbulence velocity dispersion  $\sigma_{T}$ vary in the range  $\sim 5 $ to $20$ km sec$^{-1}$. On the other hand, flattening velocity of the rotation curve for the same galaxies  has the range $\sim 100$ to $200$ km sec$^{-1}$ \citep{2008AJ....136.2648D}. Here we adopt $\sigma_{T} = f_{v} v_{0}$, with $f_{v} = 0.1$. Considering the gas at a temperature of $\sim 500$ K, we adopt the thermal velocity dispersion  to be  $\sigma = f_{T} v_{0}$, where $f_{T} = 0.01$. Note that the actual value of $v_{0}$ is unimportant here.

In order to see the effect of the density and velocity fluctuations in $P(K, \delta v)$, we consider  $\alpha = (-1.5, -2.5)$, $\beta = (-1.5, -2.5)$ and all combinations of these. In literature the velocity and density fluctuations because of turbulence are assumed to be uncorrelated. Here we consider two cases, either $n_{HI}$ and $v^{T}$ are completely uncorrelated or completely correlated. Galaxies with higher inclination angles can have scale mixing in the projected direction.  Rotational velocity effects also manifest more at higher inclination angle because of the $sin(i)$ factor in eqn.~(13). We choose the inclination angle to be $i= 10 ^{\circ}$. 

Figure~(1) shows different aspects of the  simulated galaxy for $\alpha = -2.5$, $\beta = -2.5$, $i=10^{\circ}$ and for the case when $n_{HI}$ and $v^{T}$ are uncorrelated.  Column density map is shown in Figure~(1a), while the line of sight velocity $v_{z}$ is shown in Figure~(1b). Figure~(1c) shows $I^{obs}$ for a certain value of $v$ with $\delta v = \sigma_{T}/2$. In this case only a part of the galaxy is visible and the area over which $P(K, \delta v)$ can be estimated is restricted. The integrated line profile of the galaxy is shown in Figure~(1d).

Assuming the centre of the galaxy to be at the centre of the simulation volume and the inclination angle to be $i$, we generated the specific intensity given in eqn.~(1) with velocity resolution same as thermal velocity dispersion. Range of $v$ is chosen such that all the emission from the model galaxy is included.
We  estimate $P(K, \delta v)$ defined in eqn.~(7) for  (a) $\delta v$ covering the entire \HI emission, i.e, $\delta v = \Delta v$  and (b) $\delta v = \sigma_{T}/2$. Results are discussed in the next section.

\section{Results and Discussions}

\begin{figure}
\begin{center}
\epsfig{file=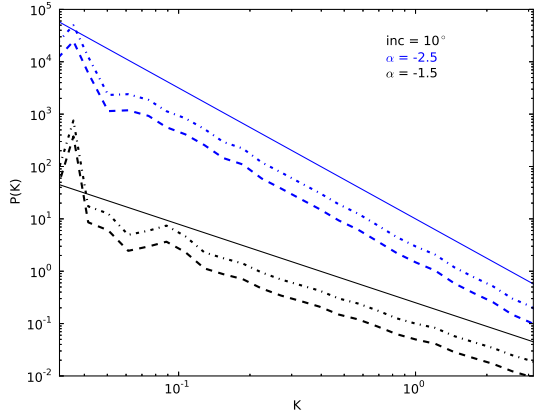,width=3.2in}
\end{center}
\caption{2D power spectra with $\delta v = \Delta v$ plotted for different combinations of $\alpha = (-1.5, -2.5)$ and $\beta = (-1.5, 2.5)$ for inclination angle $i = 10^{\circ}$. Blue dash and dot-dash lines show the power spectra for $\beta = -1.5$ and $-2.5$ respectively with $\alpha = -2.5$. The solid blue line is a plot of $P(K) \propto k^{\alpha}$.  The black curves are for corresponding power spectra with $\alpha=-1.5$. All curves are shifted arbitrarily in the vertical direction for clarity.}
\label{fig:NHI}
\end{figure}

We first discuss the results for the case when we assume that the $n_{HI}$ and $v^{T}$ are uncorrelated. Figure~(2) shows the power spectra $P(K, \delta v)$ with $\delta v = \Delta v$. Here the``dot-dash" lines corresponds to the power spectra for $\beta = -2.5$ while the  ``dash-dash" lines are for $\beta = -1.5$. Power spectra corresponding to $\alpha = -2.5$ are shown in blue and $\alpha = -1.5$ are shown in black. The solid lines corresponds to power law with slopes $-2.5$ and $-1.5$ respectively. All curves are shifted in vertical direction arbitrarily for clarity. It is clear that the power spectra of the intensity with $\delta v = \Delta v$ has the same slope that of the $n_{HI}$, irrespective of the slope of the velocity power spectra. As in our simulation we have considered a thick disk for the galaxy, we expect the slope of the power spectrum of column density to be same with that of $n_{HI}$. Hence, nature of the power spectra in Figure~(1) is quite expected and is just a verification of the eqn~(8). 

\begin{figure}
\begin{center}
\epsfig{file=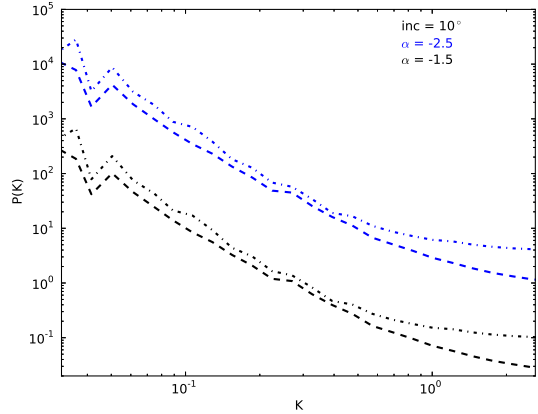,width=3.2in}
\end{center}
\caption{2D power spectra with $\delta v = \sigma_{T}/2$ plotted for different combinations  as in Figure~(1). Note that slope of the spectra changes for $K > 0.8$. }
\label{fig:NHI}
\end{figure}

We estimate the power spectra $P(K, \delta v)$ for all four combinations of $\alpha$ and $\beta$ and $i = 10^{\circ}$ with $\delta v = \sigma_{T}/2$, where we expect to see the effect of turbulence velocity $v^{T}$ in the intensity power spectra. Figure~(3) shows the corresponding power spectra as in Figure~(1) with $\delta v = \sigma_{T}/2$. Before we interpret these curves, we need to realise that here emission is coming from only a part of the galaxy's disk, as shown in Figure~(1c). In such a case, the observed intensity power spectra would have effect of  the shape of the window where the emission is coming from \footnote{Effect of the window is discussed in detail in \citet{2009MNRAS.398..887D}}. This is precisely why all four curves in Figure~(3) have similar nature for $K<0.8$ and we can only expect to see the effect of $\alpha$ or $\beta$ beyond that. Interestingly, for $K>0.8$ the power spectra is independent of the values of $\alpha$ and is different for different $\beta$. This can be the effect of velocity modification, i.e, effect of the line of sight component of the turbulent velocity $v^{T}$ on the intensity power spectrum. As this is independent of $\alpha$, the nature of the velocity modification is different than what is expected from the result of LP00 (see eqn.~(10, 11)).

To investigate how $P(K, \delta v)$  for $K>0.8$ changes with different values of $\beta$, we performed the same simulation with $\alpha = -2.5$ and $i=10^{\circ}$, for values of $\beta$ ranging $-3.0$ to $-1.0$. For each case, we fit the power spectra at $K>0.8$ with a power law of the form $P(K) \propto K^{\gamma}$ and note the best fit values. Since in simulation we have not added any contribution from the observational uncertainties, we only use the sample variance generated noise to do this fit. Results are shown in Figure~(4), where we have plotted $\beta$ in x axis with $\gamma$ with errors form the fit in y axis. We use a second order polynomial to empirically  fit the values of $\gamma$ against $\beta$, i.e, $f(x) = a_{0} + a_{1} x + a_{2} x^{2}$ with $a_{0}, a_{1}$ and $a_{2}$ values    $-0.27, -0.17$ and $-0.15$ respectively.

\begin{figure}
\begin{center}
\epsfig{file=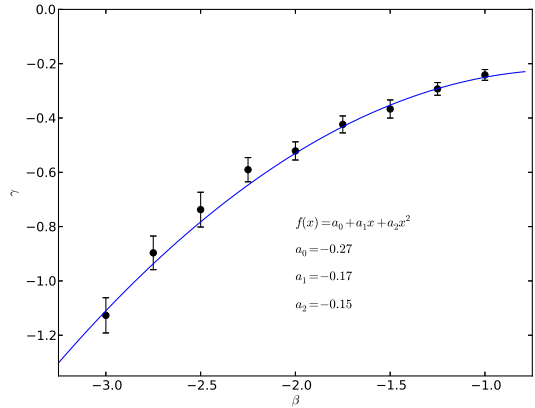,width=3.2in}
\end{center}
\caption{ Points with error bars represents best fit values of $\gamma$ as a function of $\beta$. We use a polynomial fit to the points and the best fit line is shown in blue solid line. The values of the polynomial coefficients are also given in the plot window.}
\end{figure}

Next we consider the case when the fluctuations in $n_{HI}$ and that in $v^{T}$ are correlated, in fact we use the same set of gaussian random variables to represent them. Hence, in this case, we only consider variation of $\alpha$ since  $n_{HI}$ and $v^{T}$ fluctuations are  scaled up version of the same original gaussian variables.
As expected the power spectrum with $\delta v = \Delta v$ has the slope of the column density power spectrum and the corresponding plot is exactly similar to Figure~(2) except from minute differences arising because of statistical fluctuations. We estimated the power spectrum of \HI intensity with $\delta v = \sigma^{T}/2$. These power spectra also show trends as in Figure~(3), for $K<0.8$ it is dominated by the windowing effect and for larger $K$, it is a power law with exactly similar variation of the slope of the spectra as in Figure(4). We do not show these plots here to avoid repetition.

To summarise, for both the cases $n_{HI}$ and $v^{T}$ uncorrelated and perfectly correlated the power spectrum $P(K, \delta v)$ with $\delta v = \Delta v$, we always reproduce power law with the same slope as $n_{HI}$ power spectrum and with $\delta v = \sigma_{T}/2$, at larger $K$ the power spectrum has a certain slope that  only correlates with the slope of the velocity spectrum. 

\section{Conclusions}
In this letter we investigate how the \HI intensity fluctuation power spectrum is related to the number density  and the line of sight component of the turbulent velocity for external galaxies. We found that for scale invariant fluctuations in both density and velocity, when the  emission is integrated over all velocity range, the intensity fluctuation power spectra follow a power law that has the same slope of the power spectra of the \HI number density fluctuations. We consider a thick disk for the galaxy in this case, in case of thin disk, the intensity spectra would have slope shallower by order unity (\citet{2009MNRAS.398..887D}, LP00).

When the emission is integrated over a velocity range smaller than the turbulent velocity dispersion, due to the galactic rotation, only a  part of the galaxy is visible. Effect of the density or the velocity fluctuation in the intensity power spectra can be inferred only for higher values of $K$ and for a relatively narrow range of $K$ values. We found that the slope of the spectra at these range follow approximately a power law with slope $\gamma$ related nonlinearly only to the slope of the velocity power spectrum $\beta$ and independent of the power spectrum of the density. This is a different result compared to LP00, where it is expected that $\gamma = \alpha +\beta/2$. Note that, this result is based on a power law fit to the power spectrum for a shallower range of $K$. Nevertheless, it clearly demonstrates that velocity modification of the \HI power spectrum for external galaxies is quite different from that in our Galaxy.

In our simulations, we did several simplifications. First we ignored the overall \HI profile of the galaxy. However, as it is shown in \citet{2009MNRAS.398..887D}, effect of this profile is to modify the power spectra at lower $K$, which considering the galaxy spanning over the simulation volume would happen at $K<0.1$ and is not the regime of interest for the velocity modification. Similarly, including a realistic rotation curve would have only change the window over which the power spectrum can be investigated for velocity modification. Given these, our results also stand out for real galaxies. Effect of varying thermal velocity dispersion over the galaxy, on the other hand, is more complicated and needs to be investigated in detail separately.

Finally, we discuss the feasibility to use the relation we obtain between $\gamma$ and $\beta$ in Figure~(4) for a real observation.  Considering  the galaxy spread over our entire simulation volume, the dynamic range in $K$ from simulation is  approximately same as that in the THINGS observation.  \citet{2013NewA...19...89D} has estimated the power spectra of 18 nearby spiral galaxies from THIGNS sample.  Given the baseline coverage of observation, they could estimate the power spectra till $1/4^{th}$ the largest baseline. This is because at higher baselines the baseline coverage is restricted and signal to noise is insufficient to estimate the power spectrum with statistical significance. We use $NGrid = 512$ for performing our simulation. As the largest $K$ in simulation is $\pi$, $1/4^{th}$ the available baseline rang of THINGS corresponds to $K = \pi /4 \sim 0.8$ in our simulation, making these observation inefficient to probe the velocity modification this way. We choose an inclination angle of $10^{\circ}$ for our simulation, any higher inclination angle would result rather even smaller range in $K$  over which the values of $\gamma$ can be estimated.

We conclude that with present telescopes to use VCA techniques for external galaxies, we need high integration time. An alternate method to estimate the turbulent velocity spectra of the external galaxies would be more useful. We aim to investigate along this direction in future.

\section*{Acknowledgement}
PD acknowledges useful discussion with Somnath Bharadwaj, Jayaram N. Chengalur, Nirupam Roy and Nissim Kanekar. This work is supported by the DST INSPIRE Faculty Fellowship award  [IFA-13 PH 54] and performed at Indian Institute of Science Education and Research, Bhopal.
PD is thankful to Narendra Nath Patra, Sushma Kurapati and Preetish Kumar Mishra for reading the earlier version of the draft and providing valuable comments.

\end{document}